\documentclass[aps,prb,10pt,twocolumn,a4paper,superscriptaddress,nofootinbib,longbibliography,floatfix]{revtex4-2}

\usepackage{graphicx}
\usepackage{amsmath}
\usepackage{amssymb}
\usepackage{amsfonts}
\usepackage{color}
\usepackage{xcolor}
\usepackage[colorlinks = true,
            linkcolor = blue,
            urlcolor  = blue,
            citecolor = blue,
            anchorcolor = blue,
            unicode]{hyperref}

\DeclareMathOperator{\Tr}{Tr}
\begin{document}

\title{Superconductivity-Enabled Conversion of Ferromagnetic Resonance into Standing Spin Waves}

\author{Ya.\ V.\ Turkin}
\affiliation{HSE University, Moscow 101000, Russia}

\author{N.\ G.\ Pugach}
\email{npugach@hse.ru}
\affiliation{HSE University, Moscow 101000, Russia}

\author{F.\ M.\ Maksimov}
\affiliation{Russian Quantum Center, Skolkovo, Moscow 143025, Russia}
\affiliation{Moscow Institute of Physics and Technology, Institutsky lane 9, Dolgoprudny, Moscow region, 141700}

\author{A.\ S.\ Pakhomov}
\affiliation{Russian Quantum Center, Skolkovo, Moscow 143025, Russia}
\affiliation{Moscow Institute of Physics and Technology, Institutsky lane 9, Dolgoprudny, Moscow region, 141700}

\author{S.\ N.\ Polulyakh}
\affiliation{Physical-Technical Institute, Vernadsky Crimean Federal University, Simferopol 295007, Russia}

\author{V.\ S.\ Stolyarov}
\affiliation{Centre for Advanced Mesoscience and Nanotechnology, Moscow Institute of Physics and Technology, Dolgoprudny, Moscow region 141700, Russia}

\author{A.\ I.\ Chernov}
\affiliation{Russian Quantum Center, Skolkovo, Moscow 143025, Russia}
\affiliation{Moscow Institute of Physics and Technology, Institutsky lane 9, Dolgoprudny, Moscow region, 141700}

\author{V.\ I.\ Belotelov}
\affiliation{Russian Quantum Center, Skolkovo, Moscow 143025, Russia}
\affiliation{Lomonosov Moscow State University, Leninskie Gory 1(2), Moscow 119991, Russia}

\date{\today}
\begin{abstract}
Superconductors can transport spin without Joule dissipation, yet their coherent coupling to short-wavelength magnons in insulating magnets remains largely unexplored.
Here we demonstrate experimentally and theoretically that a conventional diffusive superconductor can \emph{enable} the conversion of the uniform ferromagnetic resonance (FMR) mode into perpendicular standing spin waves (PSSWs) in an adjacent ferrimagnetic insulator.
In Bi-substituted iron-garnet/Nb bilayers, the microwave transmission develops an additional resonance feature that appears only below the Nb superconducting transition temperature and lies close to the uniform FMR peak.
A microscopic theory that self-consistently couples the quasiclassical Keldysh--Usadel description of the superconducting condensate to the Landau--Lifshitz--Gilbert dynamics shows that the conversion relies on two ingredients: (i) an interfacial spin-transfer torque mediated by spin-polarized triplet Cooper pairs and (ii) a depth-dependent effective field produced by Abrikosov vortices (electromagnetic proximity).
The resulting susceptibility reproduces the measured lineshapes and establishes superconductivity as an active control parameter for exchange standing-wave modes in magnetic insulators.
\end{abstract}

\maketitle
Superconductor/ferromagnet hybrids are a key platform for superconducting spintronics and magnonics, where superconducting correlations and collective magnetic excitations can be combined for low-dissipation information processing \cite{dobrovolskiy2026roadmap,cai2023superconductor,mel2022superconducting,yu2022efficient,li2022coherent,golovchanskiy2021ultrastrong,gusev2021magnonic,barman20212021}.
Ferromagnetic resonance (FMR) in such hybrids is a versatile probe of Meissner screening, vortex physics, and spin-transfer phenomena \cite{bell2008spin,jeon2019effect,jeon2019abrikosov,jeon2018enhanced,li2018possible,chan2023controlling,PhysRevB.97.224414,jeon2019exchange,umeda2018spin,muller2021temperature}.
To date, however, superconductivity-driven FMR effects have largely been discussed for the \emph{uniform} ($k{=}0$) mode, as renormalizations of the resonance field and Gilbert damping due to interfacial spin torques and electromagnetic screening \cite{houzet2008ferromagnetic,morten2008proximity,yokoyama2009tuning,silaev2020finite,simensen2021spin,montiel2023spin}.
In contrast, controlling \emph{exchange} magnons (finite-$k$) in insulating magnets typically requires either (i) a second magnetic layer enabling resonant spin pumping and mode hybridization \cite{klingler2018spin,qin2018exchange} or (ii) strongly nonuniform microwave fields, e.g., eddy-current shielding in conducting ferromagnets \cite{kostylev2009strong,schoen2015radiative,Kalashnikova}.
Very recently, moving Abrikosov vortex lattices were shown to provide an electrically driven nanoscale source of short-wavelength magnons, enabling unidirectional sub-40-nm spin-wave excitation via the vortices' stray fields \cite{dobrovolskiy2025moving}.
A key open question is whether a conventional superconductor---which has no intrinsic magnetic resonances at gigahertz frequencies---can nevertheless \emph{actively populate} finite-$k$ standing-wave modes in a nearby magnetic insulator.

Here we demonstrate that superconductivity can \emph{enable} efficient transfer of energy from the uniform FMR precession to perpendicular standing spin waves (PSSWs) in an adjacent ferrimagnetic insulator.
In Bi-substituted iron-garnet/Nb bilayers driven by a spatially uniform microwave field, the transmission spectrum develops an additional resonance feature that appears only below the Nb superconducting transition temperature and lies in close proximity to the uniform FMR peak.
We identify this feature as the lowest exchange PSSW activated by a superconductivity-induced \emph{mode-conversion channel} between $k{=}0$ and finite-$k$ magnons.

\begin{figure*}
  \includegraphics[width=\textwidth]{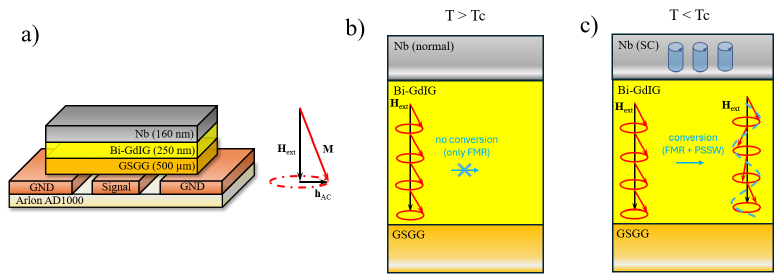}
  \caption{\label{fig:setup}
  a) Sample and measurement geometry. A Bi-substituted iron-garnet film is capped by a Nb layer and placed on a $50~\Omega$ coplanar waveguide. The microwave magnetic field $\mathbf{h}_{\mathrm{AC}}$ drives magnetization precession about the out-of-plane static field $\mathbf{H}_{\mathrm{ext}}$. b) and c) Schematic illustration of the observed effect of conversion of the FMR into PSSW under the action of vortices.
  b) When Nb is above its critical temperature, only the ferromagnetic resonance mode is observed. c) Below the critical temperature, a conversion of the FMR mode is observed with the appearance of PSSW under the influence of superconducting vortices.}
\end{figure*}

The samples consist of a Bi-substituted iron-garnet film  \texorpdfstring{$(\mathrm{BiGd})_3(\mathrm{FeSc})_5\mathrm O_{12}$}{(BiGd)3(FeSc)5O12} (Bi-GdIG, thickness $d_{\mathrm{FI}}=250$~nm) grown on a (111) gadolinium scandium gallium garnet substrate (GSGG, $d_{\mathrm{sub}}=500~\mu$m) and capped by a diffusive Nb film ($d_{\mathrm{Nb}}=160$~nm).
A reference, uncapped Bi-GdIG film from the same growth batch was measured under identical conditions.
For broadband microwave spectroscopy, the samples were mounted face-down on a $50~\Omega$ coplanar waveguide and placed in a closed-cycle cryostat between solenoidal coils such that $\mathbf{H}_{\mathrm{ext}}$ is perpendicular to the film plane (out-of-plane geometry).
The complex transmission $S_{21}(H,f)$ was recorded by a vector network analyzer while sweeping the external field at a fixed frequency.
In the data analysis, a smooth background was subtracted and the signal amplitude was normalized (see Supplemental Material \cite{supp} for the full procedure, CPW geometry, and complete temperature series).

In the out-of-plane configuration and at fields exceeding the saturation field, the uniform FMR frequency follows the Kittel relation for thin films, and its temperature dependence is dominated by the evolution of the effective magnetization and anisotropy \cite{lee2016ferromagnetic}.
A crucial advantage of this geometry is that, once the superconductor enters the mixed state, Abrikosov vortices generate a depth-dependent stray field near the FI/S interface that breaks the mirror symmetry across the film thickness and thereby couples the uniform microwave drive to standing-wave eigenfunctions.

\begin{figure*}
  \includegraphics[width=\textwidth]{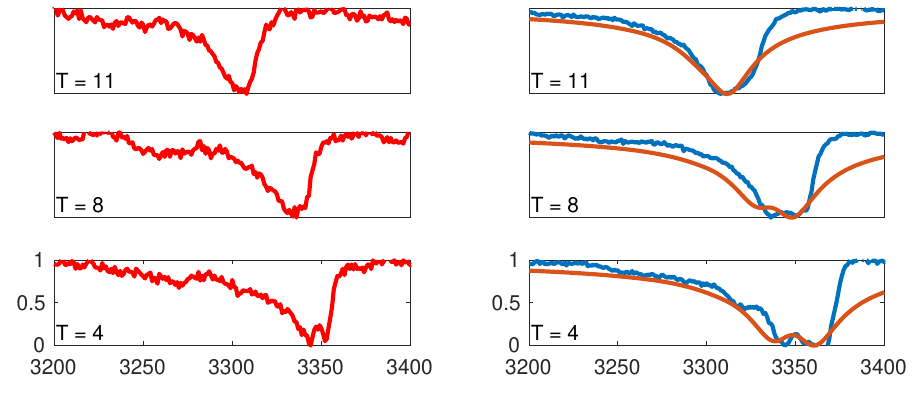}
  \caption{\label{fig:traces}
  Normalized microwave transmission $|S_{21}|$ versus out-of-plane field at $f=4$~GHz for representative temperatures $T=11$~K (normal state), $T=8$~K (near $T_c$), and $T=4$~K (deep in the superconducting state).
  Left column: uncapped Bi-GdIG film (experiment, black) together with the LLG theory (red).
  Right column: Bi-GdIG/Nb bilayer (experiment, orange) together with the computed lineshape (theory, magenta).
  Below $T_c$, the bilayer develops an additional resonance feature (the ``right'' peak) near the uniform FMR position, while the uncapped film shows only the conventional response. For more details see Supplemental Material \cite{supp}}
\end{figure*}

Figure~\ref{fig:traces} summarizes the key observation.
In the uncapped Bi-GdIG film, the response is dominated by a single FMR feature whose resonance field shifts smoothly with temperature.
At the lowest temperatures, a weak splitting is visible; we attribute it to a thin interfacial magnetic sublayer produced by substrate-element diffusion during growth \cite{lutsev2016low,lutsev2020spin}.
Notably, the uncapped film does not show a pronounced additional peak of comparable amplitude, indicating that a uniform microwave field alone does not efficiently excite PSSWs in this insulating garnet within the studied frequency window.

In striking contrast, the Bi-GdIG/Nb bilayer exhibits a strong asymmetry and a second resonance feature that emerges only in the superconducting state.
We refer to the two low-temperature features as the ``left'' and ``right'' peaks according to their field positions.
The left peak continuously connects to the reference-film FMR and therefore corresponds to the uniform mode.
The right peak appears upon cooling through $T_c$ and grows as superconductivity develops.
Its amplitude becomes comparable to that of the left peak, which is a hallmark of efficient energy transfer from the driven uniform precession to a secondary mode.
Such a prominent secondary resonance is difficult to reconcile with simple standing-wave quantization: without an additional symmetry-breaking mechanism, the overlap between a spatially uniform drive and higher-order standing-wave eigenfunctions is strongly suppressed.
The superconductivity-triggered onset therefore points to a new coupling channel introduced by the superconducting condensate.

Our interpretation is that superconductivity provides an efficient \emph{mode-conversion} pathway between the uniform FMR and exchange PSSWs.
Two mechanisms are essential and act cooperatively.

(i) \emph{Triplet-mediated interfacial torque and pinning.}
At a spin-active FI/S interface, the precessing magnetization generates spin-polarized triplet Cooper correlations in the diffusive superconductor.
These correlations produce an additional interfacial spin-transfer torque and effective dynamic interfacial pinning, which can be expressed in terms of the static and dynamic interfacial spin susceptibilities $\chi_0$ and $\chi(\Omega)$ \cite{bobkova2018spin,silaev2020finite,eschrig2015general,ouassou2017triplet}.
For a recent review of proximity-induced triplet correlations at spin-active magnetic interfaces, including N\'eel-type triplets in superconductor/antiferromagnet hybrids, see Ref.~\cite{bobkova2024neel}.
Because $\chi$ depends sensitively on the superconducting state, this torque channel is strongly modified below $T_c$.

(ii) \emph{Vortex-induced depth-dependent field (electromagnetic proximity).}
In our experiment the out-of-plane fields are well above the first critical field of Nb (so vortices form) yet far below $H_{c2}$ (so the mixed state is dilute).
Abrikosov vortices distort the static magnetic field near the FI/S interface and create a depth-dependent effective field within the ferrimagnet.
This spatial inhomogeneity breaks the mirror symmetry across the garnet thickness and enhances the overlap between the uniform microwave drive and standing-wave modes \cite{dobrovolskiy2019magnon,tinkham2004introduction}.
We emphasize that this electromagnetic proximity effect is absent in the uncapped reference film.

A further ingredient that makes the conversion particularly efficient in our system is the relatively small exchange stiffness of Bi-substituted garnets compared to standard YIG, which compresses the PSSW spectrum and brings the lowest exchange mode close to the uniform FMR frequency.
This near-degeneracy enables strong hybridization once superconductivity provides the required symmetry breaking and interfacial coupling.

Because the ferrimagnet is a good electrical insulator, the commonly invoked eddy-current mechanism for PSSW excitation in metallic films \cite{kostylev2009strong,schoen2015radiative} is inoperative here.
Likewise, the additional feature is not present in the uncapped film and thus cannot be attributed solely to the microwave-field profile of the CPW.
These observations further support the conclusion that the superconducting state of Nb provides the dominant symmetry breaking and coupling needed to excite finite-$k$ modes.

In the perpendicular geometry, thickness-quantized exchange modes obey $k_n \approx n\pi/d_{\mathrm{FI}}$ and their frequencies are shifted by the exchange term $\propto A_{\mathrm{ex}}k_n^2$ in the effective field.
For Bi-substituted garnets, the reduced exchange stiffness compresses the mode spectrum so that the lowest PSSW can approach the uniform mode within a few linewidths, making it susceptible to hybridization once an interfacial pinning gradient and a depth-dependent field are present.
Superconductivity therefore plays a dual role: it activates the coupling (through triplet proximity) and supplies the spatial asymmetry (through vortices) needed to overcome the symmetry suppression of uniform driving.

To capture the observed spectra we model the ferrimagnetic dynamics by a linearized one-dimensional Landau--Lifshitz--Gilbert equation for the transverse magnetization $\mathbf{m}(z,t)$,
\begin{equation}
\begin{aligned}
\partial_t \mathbf{m}
&= \gamma \bigl(H_0+H_{\mathrm{SC}}(z)\bigr)\,\mathbf{e}_z\times \mathbf{m}
- \frac{2A_{\mathrm{ex}}\gamma}{\mu_0 M_s}\,\mathbf{e}_z\times \partial_z^2\mathbf{m}\\
&\quad - \gamma M_s\,\mathbf{e}_z\times \mathbf{h}_{\mathrm{AC}}
+\alpha_{0}\,\mathbf{e}_{z}\times
\frac{\partial\mathbf{m}}{\partial t},
\end{aligned}
\label{eq:llg}
\end{equation}
where $A_{\mathrm{ex}}$ is the exchange stiffness, $M_s$ is the saturation magnetization, and $\alpha_0$ is the intrinsic Gilbert-damping parameter.
The superconducting contribution $H_{\mathrm{SC}}(z)$ represents the vortex-induced depth-dependent field and is modeled by a short-range profile near the interface (details and parameter values are provided in the Supplemental Material \cite{supp}).
More precisely, the component of the microscopic vortex-lattice field along the static field is spatially nonuniform both in the film plane and across the interface and can be expanded as $H_{\mathrm{v},z}(\mathbf{r}_{\parallel},z)=\sum_{\mathbf{G}}H_{\mathbf{G}}(z)e^{i\mathbf{G}\cdot\mathbf{r}_{\parallel}}$, where $\mathbf{G}$ are reciprocal-lattice vectors. In the present one-dimensional description, $H_{\mathrm{SC}}(z)$ denotes the effective profile obtained by projecting the full electromagnetic perturbation onto the in-plane mode sector retained in the ferrimagnetic model; it should therefore not be interpreted as a literal magnetic field that is uniform in the film plane and varies only with $z$.
The proximity-induced interfacial torque enters through a spin-active boundary condition at the FI/S interface,
\begin{equation}
\mathbf{e}_z \times \partial_z \mathbf{m}\big|_{z=0}
= \zeta\bigl[\chi(\Omega)-\chi_0\bigr]\;\mathbf{e}_z\times \mathbf{m}\big|_{z=0},
\label{eq:bc}
\end{equation}
which is of Robin type and encodes an effective (temperature-dependent) interfacial pinning strength. Here, $\zeta$ is a phenomenological FI/S torque-conversion coefficient.
The susceptibilities $\chi_0$ and $\chi(\Omega)$ are obtained from a microscopic quasiclassical Keldysh--Usadel description of the diffusive superconductor with spin-mixing boundary conditions.

We compute $\chi_0$ and $\chi(\Omega)$ for a diffusive Nb film using the quasiclassical Keldysh--Usadel formalism (see, e.g., Refs.~\cite{brinkman2003microscopic,larkin1977non,heikkila2019thermal}).
In the linear-response regime relevant for FMR, the spin pumping of the superconductor is described by the first-order correction $\check{g}^{(1)}_n$ to the equilibrium Green function $\check{g}^{(0)}$ at Fourier harmonic $n=\pm 1$.
The linearized Usadel equation takes the form \cite{silaev2020finite}
\begin{equation}
\begin{aligned}
\hbar D \, \check{g}^{(0)}(\omega_{n+})\partial_{z}^{2}\check{g}^{(1)}_{n}(\omega)+
\left[\hbar\omega\check{\rho}_{3},\check{g}^{(1)}_{n}(\omega)\right]
\\
+ \frac{n\hbar\Omega}{2}
\left\{\check{\rho}_{3},\check{g}^{(1)}_{n}(\omega)\right\}
= 
&i\left[\check{g}^{(1)}_{n}(\omega),\check{\Delta}^{(0)}\right],
\label{eq:usadel_lin}
\end{aligned}
\end{equation}
where $\hbar\omega$ is the quasiparticle energy, $D$ is the diffusion constant, $\check{\rho}_3 = \text{diag}\left[1,1,-1,-1\right]$ is the Nambu basis matrix, and $\omega_{n\pm}=\omega\pm n\Omega/2$.
The first-order correction satisfies the linearized normalization condition
\begin{equation}
\check{g}^{(0)}(\omega_{n+})\check{g}^{(1)}_{n}(\omega)
+\check{g}^{(1)}_{n}(\omega)\check{g}^{(0)}(\omega_{n-})=0.
\label{eq:usadel_normalization}
\end{equation}
At the spin-active FI/S interface we use a spin-mixing boundary condition parameterized by the phenomenological parameter $\kappa_{\phi}$ containing spin-mixing angle $\phi$\cite{eschrig2015general,ouassou2017triplet}:
\begin{equation}
    \begin{aligned}
\check{g}^{(0)}(\omega_{n+})\,\partial_{z}\check{g}^{(1)}_{n}(\omega)
=& i\kappa_{\phi} \left[  \check{g}^{(0)}(\omega_{n+})\check{m}_{n} \right.
\\
\left.
- \check{m}_{n}\check{g}^{(0)}(\omega_{n-})
\right],
\label{eq:usadel_bc}
\end{aligned}
\end{equation}
where
\begin{equation}
\check{m}_{n}=\check{\rho}_{3}\boldsymbol{\check{\sigma}}
\cdot\frac{\mathbf{m}_{n}}{M_s},
\qquad
\kappa_{\phi}=\frac{g_{\phi}}{2\sigma_{\mathrm{Nb}}},
\qquad
g_{\phi}=2\frac{e^2}{hA}\sum_{\nu}\phi_{\nu}.
\label{eq:kappa_phi}
\end{equation}
Here, $g_{\phi}$ is the spin-mixing conductance per unit area, $\sigma_{\mathrm{Nb}}$ is the normal-state conductivity of Nb, $A$ is the interface area, and $\phi_{\nu}$ is the spin-mixing angle of channel $\nu$.
For a spherical Fermi surface and a channel-independent spin-mixing angle $\phi$, this reduces to
\begin{equation}
\kappa_{\phi}=\frac{e^2k_F^2\phi}{4\pi h\sigma_{\mathrm{Nb}}}.
\label{eq:kappa_phi_spherical}
\end{equation}
At the outer Nb/vacuum interface we impose a zero-current condition, $\partial_z \check{g}^{(1)}_n=0$.

The induced electronic spin magnetization in the superconductor is obtained from the Keldysh component as
\begin{equation}
\mathbf{M}^{\mathrm{SC}}_{n}
=
\frac{\hbar g  \mu_B N_{0}}{8}
\int_{-\omega_D}^{\omega_D}
 d\omega\;\
\Tr_{\mathrm{N}\otimes\mathrm{s}}\!\left[
\hat{\bm{\sigma}}\,\hat{g}^{K(1)}_{n}(\omega)\right],
\label{eq:spin_magnetization}
\end{equation}
where $N_0$ is the single-spin density of states at the Fermi level, $g$ is the electron g-factor, $\mu_B$ is the Bohr magneton, $\hat{\bm{\sigma}}=\tau_3\bm{\sigma}$ is the spin-matrix vector in Nambu--spin space, and $\omega_D$ is the cutoff. The trace is taken over Nambu and spin indices.
Since $\hat{g}^{K(1)}_{n}$ depends linearly on the interfacial magnetization $\mathbf{m}_{n}$, Eq.~(\ref{eq:spin_magnetization}) defines the dynamic susceptibility through $\mathbf{M}^{\mathrm{SC}}_{n}=\chi(\Omega)\mathbf{m}_{n}$.

Solving Eqs.~(\ref{eq:llg})--(\ref{eq:bc}) yields the standing-wave eigenmodes and their coupling to a uniform drive, from which the field-swept microwave response is computed.

The calculated susceptibility map in Fig.~\ref{fig:susc} illustrates the superconductivity-enabled hybridization.
The combined effect of vortex-induced field asymmetry and proximity-induced pinning reshapes the standing-wave spectrum such that the lowest PSSW branch approaches the uniform FMR branch within the experimentally accessed frequency window.
As a result, the driven response splits into two nearby resonances corresponding to mixed uniform/PSSW character, consistent with the ``left'' and ``right'' peaks in Fig.~\ref{fig:traces}.
Importantly, neither mechanism alone reproduces the experiment:
a purely vortex-induced $H_{\mathrm{SC}}(z)$ modifies the spectrum but yields too weak a coupling to account for the comparable peak amplitudes, whereas interfacial pinning alone requires unrealistically strong coupling and shifts the mode spectrum away from the observed resonance fields \cite{supp}.
Only their combined action yields quantitative agreement with the measured temperature evolution and lineshape asymmetry, as reflected by the theory curves in Fig.~\ref{fig:traces}. As can be seen from the Fig.~\ref{fig:traces}, the theoretical model doesn't completly fit the experimental data. The asymmetry of the FMR peak can be explained by the additional magnetostatic field, induced by the Gadolinium substrate \cite{serha2024magnetic}.

The theory curves in Fig.~\ref{fig:traces} are obtained without introducing an additional magnetic subsystem: the only dynamic degrees of freedom are the magnetization of the ferrimagnet and the superconducting condensate response encoded in $\chi(\Omega)$.
Operationally, for each temperature, we compute the superconducting spin susceptibilities from the linearized Keldysh--Usadel equations in the diffusive limit and then solve the resulting boundary-value problem for the ferrimagnet to obtain the driven response.
This procedure naturally captures both the onset at $T_c$ and the evolution of the splitting as superconductivity strengthens.
A complete derivation of the boundary conditions, the microscopic expressions for $\chi(\Omega)$, and the numerical scheme (finite-difference eigenmode expansion) are provided in the Supplemental Material \cite{supp}.

\begin{figure}
  \includegraphics[width=\linewidth]{fig3}
  \caption{\label{fig:susc}
  Calculated transverse susceptibility $\chi_{xy}$ of the Bi-GdIG/Nb structure versus frequency and field.
  The dashed line indicates the uniform-mode (Kittel) dispersion.
  Proximity-enabled hybridization with the lowest PSSW produces two nearby resonant branches, consistent with the two-peak structure observed in Fig.~\ref{fig:traces}.}
\end{figure}

Finally, the mechanism suggests clear experimental control knobs.
The vortex contribution scales with the mixed-state flux density and therefore with the out-of-plane field, while the triplet-mediated torque depends on the interfacial spin-mixing and the superconducting coherence scales.
This implies that mode conversion can be tuned not only by temperature but also by magnetic field history (vortex trapping) and, potentially, by supercurrent bias in patterned superconducting films.

In summary, we observe that superconductivity in a conventional diffusive Nb film can activate and control exchange standing spin waves in an adjacent insulating ferrimagnet driven at FMR.
The effect realizes a superconductivity-enabled conversion between the uniform magnon mode and the lowest PSSW, mediated by triplet proximity correlations and vortex-induced field inhomogeneity.
Beyond the specific Bi-GdIG/Nb system studied here, the mechanism provides a general route toward tunable magnon-mode engineering and low-dissipation magnonic functionality in superconducting hybrid platforms, where superconductivity (temperature, field, or supercurrent bias) can serve as an external control knob.
\begin{acknowledgments}
YaVT and NGP acknowledge the support of the Rus-
sian Science Foundation through Grant No. 23-72-00018 (theory and data analysis, approximations).

The experimental research was supported by the Russian Science Foundation under Grants No. 23-72-30004 (cryogenic measurements), No. 24-42-02008 (fabrication of the iron garnet thin film and its characterisation) and by the Ministry of Science and Higher Education of the Russian Federation under Project No. 075-15-2025-010 (preparation of superconductive cover of the iron garnet film).
\end{acknowledgments}
\bibliographystyle{apsrev4-2}
\bibliography{references_checked}

\end{document}